\documentclass[12pt,a4paper]{article}
\def\({\c c}
\def\|{\'\i }
\def\sqr#1#2{{\vcenter{\hrule height.#2pt
     \hbox{\vrule width.#2pt height#1pt \kern#1pt
      \vrule width.#2pt} \hrule height.#2pt}}}

\begin{document}
\title{Physical intuition and time-dependent generalizations of the Coulomb and Biot-Savart laws}
\author{Nivaldo A. Lemos\\
{\small{Departamento de F\'{\i}sica}}\\
{\small{Universidade Federal Fluminense}}\\
{\small{Av. Litor\^anea s/n, Boa Viagem - CEP 24210-340}}\\
{\small{Niter\'oi - Rio de Janeiro}}\\
{\small{Brazil}}}
\maketitle

\begin{abstract}
An analysis is performed of the  role played by physical intuition in expressing the electromagnetic field in terms of its sources in the time-dependent case.
 The conclusion is that it is 
dangerous to dissociate physical intuition from the mathematical description of the phenomena.
\end{abstract}

\newpage


Intuition is an element of paramount importance for the construction of physical theories and the solution of physical problems. 
In the case of electrodynamics, the finiteness of the speed of propagation of electromagnetic influences plays a crucial role 
 in our   understanding of electromagnetic phenomena. 

The standard argument for the retarded potentials runs as follows.
Consider the contribution to the potential at point   $\bf r$ from the charges and currents in the volume element $d\tau^{\prime}$
about point ${\bf r}^{\prime}$. Since electromagnetic influences travel at speed $c$, the  potentials at $\bf r$ at time $t$ must have originated 
in the charges and currents present in $d\tau^{\prime}$  at an instant $t_r$ {\it previous} to  $t$, so providing  electromagnetic influences 
with the time interval $\, t-t_r\,$ to propagate from   $\,{\bf r}^{\prime}\,$ to $\,{\bf r}\,$. The distance 
between   $\,{\bf r}\,$ and $\,{\bf r}^{\prime}\,$ is the magnitude of the vector $\,{\bf R} = {\bf r} - {\bf r}^{\prime}$, that is,
$\, R=\vert {\bf r} - {\bf r}^{\prime}\vert$. Thus,  the retarded time $t_r$ is determined by $\, c(t-t_r)=R$, or

\begin{equation}
\label{temporetardado}
t_r = t - \frac{R}{c} = t - \frac{\vert {\bf r} - {\bf r}^{\prime}\vert}{c}
\,\,\, .
\end{equation}
Physical intuition suggests that the potentials be given in terms of their sources by the same expressions valid in electrostatics and 
magnetostatics except for the replacement of the charge and current densities by their values at the retarded time. One is led, therefore, to introduce 
the retarded potentials

\begin{equation}
\label{potenciaisretardados}
V({\bf r},t)= \frac{1}{4\pi \epsilon_0}\int \frac{\rho ({\bf r}^{\prime},t_r)}{R}\,d\tau^{\prime}  \,\,\,\,\,\, ,\,\,\,\,\,\,
{\bf A}({\bf r},t)= \frac{\mu_0}{4\pi }\int \frac{{\bf J} ({\bf r}^{\prime},t_r)}{R}\,d\tau^{\prime}
\,\,\, .
\end{equation}

Straightforward computations \cite{Griffiths1} show that the  retarded potentials
satisfy their corresponding inhomogeneous wave equations and meet the
 Lorentz condition. 
The fields are found from the  retarded potentials by means of 

\begin{equation}
\label{camposdospotenciais}
{\bf E}= -{\mbox{\boldmath $\nabla$}}V - \frac{\partial {\bf A}}{\partial t} \,\,\,\,\,\, ,\,\,\,\,\,\,
{\bf B} = {\mbox{\boldmath $\nabla$}}\times {\bf A}
\,\,\, ,
\end{equation}
and a direct calculation  \cite{Griffiths1} yields

\begin{equation}
\label{Jefimenko1}
{\bf E}({\bf r},t) = \frac{1}{4\pi \epsilon_0}\int \bigg[ \frac{\rho ({\bf r}^{\prime},t_r)}{R^2}\,{\hat{\bf R}} +
\frac{{\dot \rho} ({\bf r}^{\prime},t_r)}{cR}\,{\hat{\bf R}} -
\frac{{\dot{\bf J}} ({\bf r}^{\prime},t_r)}{c^2R}
\bigg] 
\,d\tau^{\prime}  \,\,\, ,
\end{equation}
and
\begin{equation}
\label{Jefimenko2}
{\bf B}({\bf r},t)= \frac{\mu_0}{4\pi}\int \bigg[\frac{{\bf J}({\bf r}^{\prime},t_r)}{R^2}  + \frac{{\dot {\bf J}}({\bf r}^{\prime},t_r)}{cR}\bigg]  \times {\hat{\bf R}}\,d\tau^{\prime}
\,\,\, .
\end{equation}

These equations, in which  ${\hat{\bf R}}={\bf R}/R$ and the dot means partial  derivative with respect to time,
 are time-dependent generalizations of the Coulomb and Biot-Savart laws and appear to have been first published  by Jefimenko in 1966, 
in the first edition of his textbook \cite{Jefimenko}. These equations have been shown \cite{McDonald} to be   equivalent  to other seemingly different equations derived by  Panofsky and  Phillips \cite{Panofsky}. 

It is often emphasized  \cite{Griffiths1,Griffiths2} that the same ``logic" that worked for the potentials leads to wrong answers for the fields. 
Indeed, as Griffiths \cite{Griffiths1} remarks, ``to get the retarded {\it potentials}, all you have to do is replace $t$ by $t_r$ in the electrostatic and 
magnetostatic formulas, but in the case of the {\it fields} not only is  time replaced by retarded time, but completely new terms (involving derivatives 
of $\rho$ and ${\bf J}$) appear." This state of affairs has  been called a conundrum by McDonald \cite{McDonald}. Saying, as he does,  that the 
conundrum  is resolved by radiation is hardly a satisfying explanation of why  intuition seems to have betrayed us in the case of  the fields.

Let us take a closer look at the origin of our intuition about the potentials. In the Lorentz gauge the scalar and vector potentials obey the 
inhomogeneous wave equation

\begin{equation}
\label{ondanaohomogenea2}
\nabla^2\phi ({\bf r},t) -\frac{1}{c^2}\frac{\partial^2 \phi ({\bf r},t)}{\partial t^2} = - f ({\bf r},t)
\,\,\, .
\end{equation}
Outside the sources, that is, wherever $f ({\bf r},t)
=0$, the potential $\phi$ obeys the homogeneous wave equation, and
 $\phi$ travels at speed $c$. But the potential that propagates in vacuum emanates from the sources, which leads us to believe that the  
propagation of the influence from the   cause (source) to produce the effect  (potential) takes place at speed  $c$. Thus, the state of the  
potential at the  present time must  depend on the state of the sources  at the past  instant when electromagnetic  ``information"  left them. 
This expectation  derives from the fact that, outside the sources, the  potential satisfies  the homogeneous wave equation, whose solutions are 
known to travel at speed $c$. Therefore, coherence demands that in order to apply the same physical intuition to the fields  one must search for  
equations of the form   (\ref{ondanaohomogenea2}) for ${\bf E}$ and ${\bf B}$.

From Maxwell's equations, with the help of the identity $\,{\mbox{\boldmath $\nabla$}}\times 
({\mbox{\boldmath $\nabla$}}\times{\bf A})= {\mbox{\boldmath $\nabla$}}
({\mbox{\boldmath $\nabla$}}\cdot {\bf A})- \nabla^2 {\bf A}$, one easily gets

\begin{equation}
\label{ondanaohomogeneaE}
\nabla^2{\bf E} -\frac{1}{c^2}\frac{\partial^2 {\bf E}}{\partial t^2} = \frac{{\mbox{\boldmath $\nabla$}}\rho}{\epsilon_0} + \mu_0\,{\dot {\bf J}}
\end{equation}
and

\begin{equation}
\label{ondanaohomogeneaB}
\nabla^2{\bf B} -\frac{1}{c^2}\frac{\partial^2 {\bf B}}{\partial t^2} = -\mu_0\,{\mbox{\boldmath $\nabla$}}\times {\bf J}
\,\,\, .
\end{equation}
Now the same heuristic  argument invoked  to justify the retarded potentials suggests that

\begin{equation}
\label{IntuicaoE}
{\bf E}({\bf r},t) = -\frac{1}{4\pi \epsilon_0}\int \frac{({\mbox{\boldmath $\nabla$}}\rho ) ({\bf r}^{\prime},t_r)}{R}\, d\tau^{\prime}  -\frac{\mu_0}{4\pi}\int \frac{{\dot{\bf J}} ({\bf r}^{\prime},t_r)}{R}\, d\tau^{\prime}
 \,\,\, ,
\end{equation}
and

\begin{equation}
\label{IntuicaoB}
{\bf B}({\bf r},t) = \frac{\mu_0}{4\pi}\int \frac{( {\mbox{\boldmath $\nabla$}}\times {\bf J}) ({\bf r}^{\prime},t_r)}{R}\, d\tau^{\prime}
 \,\,\, .
\end{equation}
One must be careful with the notation:  
$\, ({\mbox{\boldmath $\nabla$}}\rho ) ({\bf r}^{\prime},t_r)\,$
denotes the  gradient of $\,\rho({\bf r},t)$, calculated keeping $t$ fixed, evaluated at $\, {\bf r}={\bf r}^{\prime}\,$ and $\, t = t_r\,$; the same goes for 
$\,( {\mbox{\boldmath $\nabla$}}\times {\bf J}) ({\bf r}^{\prime},t_r)$.

The above  expressions for $\,{\bf E}\,$ and $\,{\bf B}\,$ are not new \cite{Lorrain}. It is not obvious that these fields satisfy all 
of Maxwell's equations, since they are solutions to second order equations, whereas Maxwell's equations are of the first order. In
Lorrain and Corson \cite{Lorrain} the
 proof that these fields coincide
 with those obtained from the retarded potentials  is left to the reader, who   is asked to neglect retardation. Here we show directly that
 equations (\ref{IntuicaoE}) and (\ref{IntuicaoB}) are completely equivalent to Jefimenko's equations (\ref{Jefimenko1}) and (\ref{Jefimenko2}).

Consider the   gradient of $\,\rho({\bf r}^{\prime},t_r)\,$ with respect to $\, {\bf r}^{\prime}\,$ but now taking into account both the 
 explicit and the implicit dependences:

\begin{equation}
\label{gradientetotal}
{\mbox{\boldmath $\nabla$}}^{\prime}  \rho({\bf r}^{\prime},t_r) =
({\mbox{\boldmath $\nabla$}}\rho ) ({\bf r}^{\prime},t_r) 
+ \frac{\partial \rho({\bf r}^{\prime},t_r) }{\partial t_r}
{\mbox{\boldmath $\nabla$}}^{\prime}t_r = ({\mbox{\boldmath $\nabla$}}\rho ) ({\bf r}^{\prime},t_r) 
+\frac{ {\dot \rho}({\bf r}^{\prime},t_r) }{c}\,{\hat{\bf R}}
\,\,\, ,
\end{equation}
where we used $\, {\mbox{\boldmath $\nabla$}}^{\prime}R = - {\hat{\bf R}}\,$ and
$\, \partial t_r/\partial t =1$. Making use of (\ref{gradientetotal}) and recalling that 
 $\, \mu_0\epsilon_0 = 1/c^2$,  equation  (\ref{IntuicaoE}) can be recast as

\begin{equation}
\label{IntuicaoEquase}
{\bf E}({\bf r},t) = \frac{1}{4\pi \epsilon_0}\int \bigg[-\frac{{\mbox{\boldmath $\nabla$}}^{\prime}\rho ({\bf r}^{\prime},t_r)}{R} + \frac{{\dot \rho}({\bf r}^{\prime},t_r){\hat{\bf R}}}{cR}
-\frac{{\dot{\bf J}} ({\bf r}^{\prime},t_r)}{c^2R}\bigg]\, d\tau^{\prime}
 \,\,\, ,
\end{equation}
But 

\begin{displaymath}
\int \frac{{\mbox{\boldmath $\nabla$}}^{\prime}\rho ({\bf r}^{\prime},t_r)}{R}\, d\tau^{\prime}= 
\int\bigg[{\mbox{\boldmath $\nabla$}}^{\prime} \bigg(\frac{\rho ({\bf r}^{\prime},t_r)}{R}\bigg) - \rho ({\bf r}^{\prime},t_r)
{\mbox{\boldmath $\nabla$}}^{\prime} \bigg(\frac{1}{R}\bigg) \bigg] \, d\tau^{\prime}
\end{displaymath}

\begin{equation}
\label{integracaoporpartes}
=  \oint_{S_{\infty}}\frac{\rho ({\bf r}^{\prime},t_r)}{R} \,d{\bf a}^{\prime}
   -\int \rho ({\bf r}^{\prime},t_r)\frac{{\hat{\bf R}}}{R^2}\,d\tau^{\prime} = -\int \rho ({\bf r}^{\prime},t_r) \frac{{\hat{\bf R}}}{R^2}\,d\tau^{\prime}
\end{equation}
for localized sources (we have taken advantage of the integral theorem $\, \int_{\cal V}{\mbox{\boldmath $\nabla$}} T\, d\tau = \oint_S T\, d{\bf a}\,$ and have denoted by  $\, S_{\infty}\,$ the surface of a sphere at  infinity). With the above result,  equation
(\ref{IntuicaoEquase}) becomes identical to Jefimenko's equation  (\ref{Jefimenko1}).

Similarly, 

\begin{equation}
\label{rotacionaltotal}
{\mbox{\boldmath $\nabla$}}^{\prime}\times  {\bf J}({\bf r}^{\prime},t_r) =
({\mbox{\boldmath $\nabla$}}\times  {\bf J} ) ({\bf r}^{\prime},t_r) 
+ {\mbox{\boldmath $\nabla$}}^{\prime}t_r \times \frac{\partial {\bf J}({\bf r}^{\prime},t_r) }{\partial t_r}
 = ({\mbox{\boldmath $\nabla$}}\times  {\bf J} ) ({\bf r}^{\prime},t_r) - \frac{1}{c}\, {\dot {\bf J}}({\bf r}^{\prime},t_r) \times {\hat{\bf R}}
\,\,\, ,
\end{equation}
and (\ref{IntuicaoB}) takes the form

\begin{equation}
\label{IntuicaoBquase}
{\bf B}({\bf r},t) = \frac{\mu_0}{4\pi}\int \bigg[ \frac{{\mbox{\boldmath $\nabla$}}^{\prime}\times {\bf J} ({\bf r}^{\prime},t_r)}{R} +
\frac{ {\dot {\bf J}}({\bf r}^{\prime},t_r) \times {\hat{\bf R}}}{cR}
  \bigg]\, d\tau^{\prime}
 \,\,\, .
\end{equation}
Making an integration  by parts with the help of 

\begin{equation}
\label{Identidarotacionallinha}
{\mbox{\boldmath $\nabla$}}^{\prime} \times \bigg(\frac{{\bf J} ({\bf r}^{\prime},t_r)}{R}\bigg) =
\frac{{\mbox{\boldmath $\nabla$}}^{\prime}\times {\bf J} ({\bf r}^{\prime},t_r)}{R}
- {\bf J} ({\bf r}^{\prime},t_r)\times
{\mbox{\boldmath $\nabla$}}^{\prime} \bigg(\frac{1}{R}\bigg)
=\frac{{\mbox{\boldmath $\nabla$}}^{\prime}\times {\bf J} ({\bf r}^{\prime},t_r)}{R}
- {\bf J} ({\bf r}^{\prime},t_r)\times
\frac{\hat{\bf R}}{R^2}
\end{equation}
and dropping  the surface integral that arises from the use of the integral  theorem$\, \int_{\cal V}{\mbox{\boldmath $\nabla$}}\times {\bf A}\, d\tau = -\oint_S {\bf A} \times d{\bf a}\,$, 
one finds that equation (\ref{IntuicaoBquase}) reduces to Jefimenko's equation  (\ref{Jefimenko2}).

 Physical intuition has not led us astray, after all.
In spite of its fallibility, physical intuition is invaluable in the investigation of physical phenomena. The situation here discussed reveals, 
however, that the appeal
to intuitive arguments  requires  caution.
In particular, it is not  possible or, at least, it is dangerous to dissociate the physical intuition  from the   mathematical description of the phenomena.

\vspace{2cm}

\noindent{\bf ACKNOWLEDGMENT}

\vspace{.5cm}

The author is thankful to Jorge Sim\~oes de S\'a Martins for a critical reading of the manuscript.

\end{document}